\documentclass[a4paper,twocolumn,11pt,accepted=2017-05-09]{quantumarticle}
\pdfoutput=1
\usepackage[utf8]{inputenc}
\usepackage[english]{babel}
\usepackage[T1]{fontenc}
\usepackage{amsmath}
\usepackage{hyperref}

\usepackage{tikz}
\usepackage{lipsum}

\begin{document}

\title{Correlated Rydberg Electromagnetically Induced Transparencys}

\author{Lei Huang}
\affiliation{School of Physics and Electronic Engineering, Hainan Normal University,
Haikou 571158, P. R. China}
\orcid{0000-0002-2445-2701}
\author{Peng-fei Wang}
\affiliation{School of Physics and Electronic Engineering, Hainan Normal University,
Haikou 571158, P. R. China}
\author{Han-xiao Zhang}
\affiliation{School of Physics and Electronic Engineering, Hainan Normal University,
Haikou 571158, P. R. China}
\orcid{0000-0003-1985-4623}
\author{Yu Zhu}
\affiliation{School of Physics and Electronic Engineering, Hainan Normal University,
Haikou 571158, P. R. China}
\author{Hong Yang}
\affiliation{School of Physics and Electronic Engineering, Hainan Normal University,
Haikou 571158, P. R. China}
\orcid{0000-0003-1533-8015}
\author{Dong Yan}
\affiliation{School of Physics and Electronic Engineering, Hainan Normal University,
Haikou 571158, P. R. China}
\email{latex@quantum-journal.org}
\homepage{http://quantum-journal.org}
\orcid{0000-0003-0290-4698}
\maketitle

\begin{abstract}
  In the regime of Rydberg electromagnetically induced transparency, we study the correlated behaviors between the transmission spectra of a pair of probe fields
passing through respective parallel one-dimensional cold Rydberg ensembles. Due to the van der Waals (vdW) interactions between Rydberg atoms, each ensemble exhibits a local optical nonlinearity, where the output EIT spectra are sensitive to both the input probe intensity and the photonic statistics. More interestingly, a nonlocal optical nonlinearity emerges between two spatially separated ensembles, as the probe transmissivity and probe correlation at the exit of one Rydberg ensemble can be manipulated by the probe field at the input of the other Rydberg ensemble. Realizing correlated Rydberg EITs holds great potential for applications in quantum control, quantum network, quantum walk and so on.
\end{abstract}

In the \texttt{twocolumn} layout and without the \texttt{titlepage} option a paragraph without a previous section title may directly follow the abstract.
In \texttt{onecolumn} format or with a dedicated \texttt{titlepage}, this should be avoided.

\section{\protect\bigskip}
Rydberg atoms, which are neutral atoms in a state of high principal quantum number, are often called big atoms with exaggerated physical properties~\cite{Gallagher1}. These unusual properties result from the large orbit radius of Rydberg atoms, including long radiative
lifetimes, high polarizability and large electric dipole moments. Due to their high polarizability and large electric dipole moments, Rydberg atoms strongly interact with other Rydberg atoms~\cite{Gallagher2,Comparat} and are extremely sensitive to the surrounding electric fields~\cite{Saffman}. Undoubtedly, these features make them the natural candidates for studying many-body physics~\cite{Labuhn,Bernien,Guardado,Browaeys} and for precise measurement~\cite{Sedlacek,Anderson,Facon,Cox,Liao,Ding,Cui}.

In addition, the effect of electromagnetically induced transparency (EIT)~\cite{Harris}, as is well known in the field of quantum optics, could allow for an effective quantum interface between atoms and light without absorption. In general, photons do not directly interact with each other. However, by employing the EIT technique, the strong interactions between Rydberg atoms can be mapped onto photons, causing photons to become either strongly attractive or repulsive~\cite{Gorshkov1,Liang}. Based on the modification of photonic statistics, the combination of EIT with Rydberg atoms allows us to investigate nonlinear quantum optics at the single-photon level~\cite%
{Firstenberg} and explore quantum information applications, such as building single photon sources~\cite{Walker}, quantum gate~\cite{Muller}, transistors~\cite{Gorniaczyk,Iiarks}, filters~\cite{Peyronel}, subtractors~\cite{Honer,Gorshkov2}, and switches~\cite{Chen,Baur}.

Unlike typical linear EIT realized in an ensemble of independent atoms, Rydberg EIT spectra of the transmitted probe intensity can be influenced by the dipole blockade effect, where the excitation of two or more atoms into a Rydberg state within a mesoscopic volume is forbidden due to the dipole-dipole interaction. Specifically, the transmission coefficient and the photonic correlations become highly sensitive to the input probe intensity. Theoretical and experimental investigations on Rydberg EIT have recently attracted intense interest~\cite{Weatherill,Pritchard1,Petrosyan,Pritchard2,Reslen,Ates,Yan1,Yan2,Firstenberg2,Jen, Garttner1,Stanojevic,Li,Liu1,Liu2,Tresp,Liu3,Yan3,Tebben,Su,Srakaew,Ou}. To date, most investigations on Rydberg EIT have focused on one-dimensional systems, while studies on two-dimensional systems—specifically, to the best of our knowledge, studies on correlated Rydberg EITs—remain quite rare.

In this paper, we investigate the correlated optical responses of two probe fields passing through closely spaced, parallel one-dimensional samples of cold Rydberg atoms in the EIT regime. Each EIT spectrum exhibits cooperative optical nonlinearities when the input probe intensity is strong enough. Moreover, by varying the input probe intensity of one probe field and keeping other parameters unchanged, we observe alterations in both the transmitted probe intensity and the second-order correlation function of the other probe field. Additionally, we thoroughly examine the extent to which one probe field is influenced by changing the other probe field. The realization of correlated Rydberg
EITs enables quantum manipulation, the construction of quantum networks, the testing of quantum walk, and more.

\begin{figure}[tbph]
\includegraphics[width=0.5\textwidth,height=0.25\textwidth]{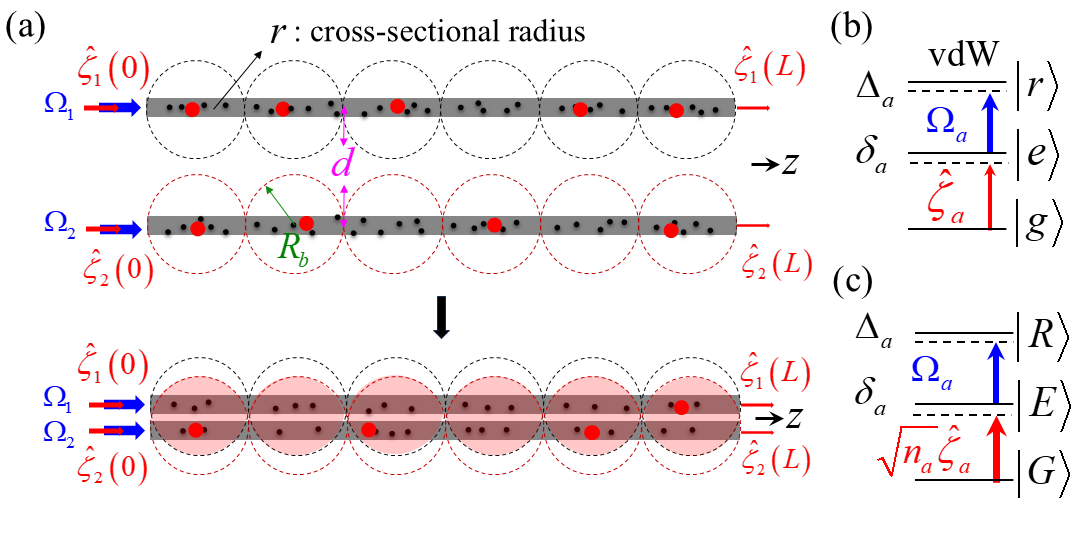}
\caption{(Color online) (a) Upper: Two weak laser fields, $\hat{\zeta}_{1}$ and $\hat{\zeta}_{2}$, propagate through two independent, parallel, one-dimensional atomic ensembles in the presence of the classical control fields $\Omega_{1}$ and $\Omega_{2}$. In this case, $d \gg R_{b} \gg r$. Lower: The optical responses of the two weak laser fields are correlated by the vdW interactions when $R_{b} \gg d \sim r$. Throughout this paper, we use the values $R_b=14.68$ $\mu \text{m}$, $d=0.5$ $\mu \text{m}$, and $r=0.5$ $\mu \text{m}$ for the numerical calculations. A rugby-shaped shaded region forms a shared blockade area, created by the overlap of two blockade spheres. (b) Atomic levels. A weak probe field (Rabi frequency operator $\hat{\zeta}_{a}$ and detuning $\delta_{a}$) and a classical coupling field (Rabi
frequency $\Omega _{a}$ and detuning $\Delta_{a}$) couple the ground state $%
|g\rangle $, intermediate state $|e\rangle $ and Rydberg state $|r\rangle $%
, respectively. $\text{vdW}$ represents the long-range van der Waals
interaction. (c) A superatom is composed of
three collective states $|G\rangle $, $|E\rangle $ and $|D\rangle $. In comparison to the single-atom case, the
collective coupling strength between states $|G\rangle$ and $|E\rangle$ is
increased by a factor of $\protect\sqrt{n_{a}}$ (a=1,2).}
\label{fig1}
\end{figure}

\section{Model and Equations}
As shown in Fig.\ref{fig1}, we consider two parallel one-dimensional $%
^{87}$Rb ultracold atomic samples, separated by the distance $d,$ both having the same cross-sectional radius $r$ and length $L$. In $a$-th $%
\left( a=1,2\right) $ atomic sample, a weak laser field $\hat{\zeta}_{a}=g%
\hat{\mathcal{E}}_{a}$ with $g$ the single atom coupling constant~\cite{Scully} and detuning $\delta _{a}$ propagates in the atomic sample in the
presence of a classical control field $\Omega _{a}$ with detuning $\Delta
_{a}$. The level scheme of the ensemble atoms is shown in Fig.\ref{fig1}(b),
$|g\rangle $, $|e\rangle $ and $|r\rangle $ are the ground state, the
excited state, and the highly excited Rydberg state of $^{87}$Rb atoms,
respectively. Specifically, these states refer to $|g\rangle
=|5S_{1/2},F=1\rangle $, $|e\rangle =|5P_{3/2}\rangle $ and $|r\rangle
=|90S\rangle $. The classical control field drives the upper transition $%
|e\rangle \rightarrow \,|r\rangle $, while the weak laser field couples the
lower transition $\,|g\rangle \rightarrow \,|e\rangle $. Together, they drive the
Rydberg atom into the three-level ladder-type configuration.

When an atom
in the $a$-th atomic sample located at $z_{ia}$ and an atom in the $b$-th atomic
sample located at $z_{jb}$ are excited to the Rydberg states, they
experience strong long-range van der Waals (vdW) interactions, where $
V_{ia,jb}=C_{6}/R_{ia,jb}^{6}$, with $C_{6}$ being the vdW coefficient and $%
R_{ia,jb}=\left\vert z_{ia}-z_{jb}\right\vert $ representing the distance between atoms.
For $a=b$, the condition $i\neq j$ holds when considering interactions between different atoms within the same sample.

 The Hamiltonian of the total system reads (%
$\hbar \equiv 1$)
\begin{equation}
\hat{H}=\sum_{a=1}^{2}\left( \hat{H}_{a}+\hat{V}_{aa}\right) +\hat{V}_{ab},
\end{equation}%
where $\hat{H}_{a}=\sum_{j}^{N}[\delta _{a}\hat{\sigma}_{ja}^{ee}+\left(
\delta _{a}+\Delta _{a}\right) \hat{\sigma}_{ja}^{rr}]+[\hat{\zeta}_{a}\hat{%
\sigma}_{ja}^{eg}+\Omega _{a}\hat{\sigma}_{ja}^{er}+$H.c$]$ is the
atom-light coupling in $a$-th atomic sample ($a=1,2$). $\hat{V}_{ab}=%
\sum_{i>j}V_{ia,jb}\hat{\sigma}_{ia}^{rr}\hat{\sigma}_{jb}^{rr}$ represents
the vdW interaction between two atoms within the $a$-th atomic sample (when $a=b$) and the vdW interaction between two atoms in the $a$ and $b$ atomic samples (when $a\neq b$). Obviously, $\hat{V}_{ab}$ vanishes, while $\hat{V}_{aa}$ still exists in the limit of $d\rightarrow
\infty $ (see upper schematic diagram in Fig.1(a)).

The dynamics of our system is governed by the following master equation
\begin{equation}
\dot{\varrho}=-i[\hat{H},\varrho ]+{\mathcal{L}}\left[ \varrho \right] ,
\label{MasterEqu}
\end{equation}%
where $\varrho $ and ${\mathcal{L}}\left[ \varrho \right] $ are the density
operator of the many-body system and Lindblad term, which accounts for the incoherent processes, respectively. To solve the many-body equation (\ref{MasterEqu}),
we can resort to the mean-field approximation. In the mean-field
description, the many-body operators are replaced by their mean values and, consequently, the interparticle correlations are completely neglected. After defining
the average atomic operator $\hat{\sigma}_{a}^{mn}\left( z\right)
=\sum_{j=1}^{n}\hat{\sigma}_{ja}^{mn}\left( z\right) /n$ in the microvolume $%
\delta V$ centered at $z$, the Heisenberg-Langevin equations for light and
atomic operators in $a$-th atomic sample read~\cite{Petrosyan}
\begin{eqnarray}
\partial _{t}\hat{\mathcal{E}}_{a}\left( z\right) &=&-c\partial _{z}\hat{%
\mathcal{E}_{a}}\left( z\right) +i\eta N\hat{\sigma}_{a}^{ge}\left( z\right)
,  \notag \\
\partial _{t}\hat{\sigma}_{a}^{ge}\left( z\right) &=&-\left( i\delta
_{a}+\gamma \right) \hat{\sigma}_{a}^{ge}\left( z\right) -ig\hat{\mathcal{E}}%
_{a}^{\dagger }\left( z\right) -i\Omega _{a}\hat{\sigma}_{a}^{gr}\left(
z\right) ,  \notag \\
\partial _{t}\hat{\sigma}_{a}^{gr}\left( z\right) &=&-i\left[ \delta
_{a}+\Delta _{a}+\hat{S}_{aa}\left( z\right) +\hat{S}_{ab}\left( z\right) %
\right] \hat{\sigma}_{a}^{gr}\left( z\right)  \notag \\
&&-\Gamma \hat{\sigma}_{a}^{gr}\left( z\right) -i\Omega _{a}\hat{\sigma}%
_{a}^{ge}\left( z\right) ,  \label{HLEqu}
\end{eqnarray}%
where $\gamma $ and $\Gamma $ are the dephasing rate in state $\left\vert
e\right\rangle $ and $\left\vert r\right\rangle ,$ respectively. $\hat{S}%
_{aa}\left( z\right) =\int d^{3}z_{a}^{\prime }\rho \left( z_{a}^{\prime
}\right) C_{6}/\left\vert z_{a}-z_{a}^{\prime }\right\vert ^{6}\hat{\sigma}%
_{a}^{rr}\left( z_{a}^{\prime }\right) $ and $\hat{S}_{ab}\left( z\right)
=\int d^{3}z_{b}^{\prime }\rho \left( z_{b}^{\prime }\right)
C_{6}/\left\vert z_{a}-z_{b}^{\prime }\right\vert ^{6}\hat{\sigma}%
_{b}^{rr}\left( z_{b}^{\prime }\right) $ denote the interaction energy shift
induced by inner Rydberg atoms in $a$-th atomic sample and the interaction energy shift induced by both a Rydberg atom in $a$-th atomic sample and all the Rydberg atoms in $b$%
-th atomic sample, respectively. Here, $\hat{S}_{aa}\left( z\right) $ and $%
\hat{S}_{ab}\left( z\right) $ have been translated into the frequency shift
like $\delta _{a}+\Delta _{a}$. It means that the Rydberg transitions from $%
|g\rangle \rightarrow \,|r\rangle $ in $a$-th atomic sample are affected not only by the atoms within the sample but also the atoms in neighbor atomic sample. In
general, both $\hat{S}_{aa}$ and $\hat{S}_{ab}$ are nonlocal in the sense
that these quantities directly depend on the density of the atomic gas and Rydberg state population.

To reasonably estimate the effects induced by the frequency shifts $\hat{S}%
_{aa}\left( z\right) $ and $\hat{S}_{ab}\left( z\right) $, we use the
superatom model, where all atoms share at most one Rydberg excitation in a
blockade region. For simplicity, we first consider the case where $\hat{S}%
_{aa}\left( z\right) \neq 0$ while $\hat{S}_{ab}\left( z\right)=0 $. As shown the upper schematic in Fig.\ref{fig1}(a), a Rydberg
superatom (SA) can generally be regarded as a sphere with the blockade radius $%
R_{b}$ due to the homogeneity and the isotropy. In the weak-probe limit,
each independent SA has three collective states $\left\vert
G_{a}\right\rangle =\left\vert g\right\rangle ^{\otimes n_{a}}$, $\left\vert
E_{a}\right\rangle =\sum_{j=1}^{n_{a}}\left\vert
g_{1},...,e_{j},...,g_{n_{a}}\right\rangle /\sqrt{n_{a}}$ and $\left\vert
R_{a}\right\rangle =\sum_{j=1}^{n_{a}}\left\vert
g_{1},...,r_{j},...,g_{n_{a}}\right\rangle /\sqrt{n_{a}}$, where $n_{a}$ is the number of atoms in the SA. These states form a three-level system as shown in Fig.1(c). Accordingly, we define the Rydberg
SA excitation projection operator as $\hat{P}_{a}=\left\vert
R_{a}\right\rangle \left\langle R_{a}\right\vert$.

Solving the Heisenberg-Langevin equations of independent SAs in the steady state, we can obtain the Rydberg excitation
projection operator~\cite{Petrosyan},
\begin{equation}
\hat{P}_{a}\left( z\right) =\frac{n_{a}\eta ^{2}\hat{\mathcal{E}}_{a}^{\dag
}\left( z\right) \hat{\mathcal{E}}_{a}\left( z\right) \Omega _{a}^{2}}{%
n_{a}\eta ^{2}\hat{\mathcal{E}}_{a}^{\dag }\left( z\right) \hat{\mathcal{E}}%
_{a}\left( z\right) \Omega _{a}^{2}+\left[ \Omega _{a}^{2}-\delta _{a}\left(
\delta _{a}+\Delta _{a}\right) \delta _{a}\right] ^{2}+\delta _{a}^{2}\gamma
_{a}^{2}}.  \label{Sdd}
\end{equation}%
The polarizability of each probe field is conditioned upon its projection,
\begin{equation}
\hat{\alpha}_{a}\left( z\right) =\hat{P}_{a}\left( z\right) \alpha _{\mathrm{TLA}}+%
\left[ \hat{1}-\hat{P}_{a}\left( z\right) \right] \alpha _{\mathrm{TLL}}  \label{ConP}
\end{equation}%
with the polarizability of a two-level atom
\begin{equation}
\alpha _{\mathrm{TLA}}=\frac{i\gamma }{\gamma +i\delta _{a}}  \label{P2}
\end{equation}%
and that of a three-level ladder atom%
\begin{equation}
\alpha _{\mathrm{TLL}}=\frac{i\gamma }{\gamma +i\delta _{a}+\frac{\Omega _{a}^{2}}{%
\Gamma +i\left( \delta _{a}+\Delta _{a}\right) }}.  \label{P3}
\end{equation}%
It is clear that the optical response of a SA depends on the
Rydberg projection operator (\ref{Sdd}): it behaves like a
two-level, absorptive medium when $\hat{\alpha}\left( z\right) $ reduces
to $\alpha _{\mathrm{TLA}}$ for $\hat{P}_{a}\left( z\right) =\hat{1}$. Alternatively, it behaves like a
three-level, EIT medium when $\hat{\alpha}\left( z\right) $ reduces
to $\alpha _{\mathrm{TLL}}$ for $\hat{P}_{a}\left( z\right) =\hat{0}$.

The transmission of the each probe light is examined through the probe light
intensity, defined as $I_{a}\left( z\right) =\langle \hat{\mathcal{E}}_{a}^{\dag }(z)%
\hat{\mathcal{E}}_{a}(z)\rangle $. The atomic sample is no longer homogeneous, and in the steady state, the propagation equation of the intensity $%
I_{a}\left( z\right) $ follows
\begin{equation}
\partial _{z}\langle \hat{\mathcal{E}}_{a}^{\dag }(z)\hat{\mathcal{E}}%
_{a}(z)\rangle =-\kappa (z)\langle \mathrm{Im}[\hat{\alpha}\left( z\right) ]%
\hat{\mathcal{E}}_{a}^{\dag }(z)\hat{\mathcal{E}}_{a}(z)\rangle ,
\label{PorEq1}
\end{equation}%
where $\kappa (z)=\rho \left( z\right) \omega _{p}/\left( \epsilon
_{0}c\gamma \right) $ is the resonant absorption coefficient. The
modification of the probe light intensity $I_{a}\left( z\right) $ is
strongly dependent on the polarizability. If $\mathrm{Im}[\hat{\alpha}%
\left( z\right) ]=0,$ i.e., under the ideal EIT condition, the probe light
intensity $I_{a}\left( z\right) $ remains unchanged in the EIT window. In general, it
decays along the $z$-axis due to $\mathrm{Im}[\hat{\alpha}\left( z\right)
]>0 $.

Next, we remove $\mathrm{Im}[\hat{\alpha}\left( z\right) ]$ out of $%
\langle \mathrm{Im}[\hat{\alpha}\left( z\right) ]\hat{\mathcal{E}}_{a}^{\dag
}(z)\hat{\mathcal{E}}_{a}(z)\rangle $ in Eq.~(\ref{PorEq1}) and simultaneously
replace $\hat{\mathcal{E}}_{a}^{\dag }(z)\hat{\mathcal{E}}_{a}(z)$ with $%
\langle \hat{\mathcal{E}}_{a}^{\dag }(z)\hat{\mathcal{E}}_{a}(z)\rangle
g_{a}\left( z\right) $ in Eq.(\ref{Sdd}), by introducing the two-photon correlation
function $g_{a}\left( z\right) =$ $\langle \hat{\mathcal{E}}_{a}^{\dag 2}(z)%
\hat{\mathcal{E}}_{a}^{2}(z)\rangle /\langle \hat{\mathcal{E}}_{a}^{\dag }(z)%
\hat{\mathcal{E}}_{a}(z)\rangle ^{2}$ under the mean-field approximation.

Similarly, the propagation equation of two-photon correlation function $%
g_{a}\left( z\right) $ follows~\cite{Petrosyan}
\begin{equation}
\partial _{z}g_{a}(z)=-\kappa (z)\hat{P}_{a}\left( z\right) \mathrm{Im}[\alpha
_{\mathrm{TLA}}-\alpha _{\mathrm{TLL}}]g_{a}(z)  \label{PorEq2}
\end{equation}%
Compared with probe light intensity $I_{a}\left( z\right) ,$ there's an additional possibility for two-photon correlation function $g_{a}\left( z\right)$
: it can be amplified by the atomic medium when $\mathrm{Im}[\alpha
_{\mathrm{TLA}}]<\mathrm{Im}[\alpha _{\mathrm{TLL}}].$

When the two atomic
samples are close but not coincident with $R_{b}\gg d\sim r$, both $\hat{S}_{aa}$ and $\hat{S}_{ab}$ are active in this regime. A single Rydberg excitation in the rugby-shaped shaded region (the mutual
blockade region) can suppress further excitations not only
within the atomic sample itself but also in the neighboring sample. Based on this, we employ a stochastic procedure to solve the coupled Eq.~(\ref%
{Sdd})-(\ref{PorEq2}) with initial input probe light intensity $I_{a}\left(
0\right) $ and its initial two-photon correlation function $g_{a}\left(
0\right) $. We divide each sample into $N=L/\left( 2R_{b}\right) $ and then simultaneously assess the Rydberg excitations of two SAs
but in the same \textit{rugby}. Specifically, $P_{a}\left( z\right) $ and $%
P_{b}\left( z\right) $ are calculated from Eq.(\ref{Sdd}) and compare them
with the respective random number $p_{a}$ and $p_{b}$, generated
independently. There are three main cases: (I) If $P_{a}\left( z\right)
<p_{a}$ and $P_{b}\left( z\right) <p_{b},$ set $P_{a}\left( z\right)
=P_{b}\left( z\right) =0;$ (II) If $P_{a}\left( z\right) \geq p_{a}$ and $P_{b}\left( z\right) <p_{b}$, set $P_{a}\left( z\right) =1$ and $%
P_{b}\left( z\right) =0$ (or vice verse)$; $(III) If $P_{a}\left( z\right) \geq p_{a}$ and $%
P_{b}\left( z\right) \geq p_{b},$ then if $P_{a}\left( z\right) >P_{b}\left(
z\right) $, set $%
P_{a}\left( z\right) =1$ and $P_{b}\left( z\right) =0$ (or vice versa); otherwise, set $%
P_{a}\left( z\right) =P_{b}\left( z\right) =0.5.$
This evaluation is carried out one-by-one using Monte Carlo sampling. To obtain steady mean values,
this procedure is typically repeated many times.

\section{Results and discussion}
\begin{figure}[tbph]
\includegraphics[width=0.5\textwidth,height=0.25\textwidth]{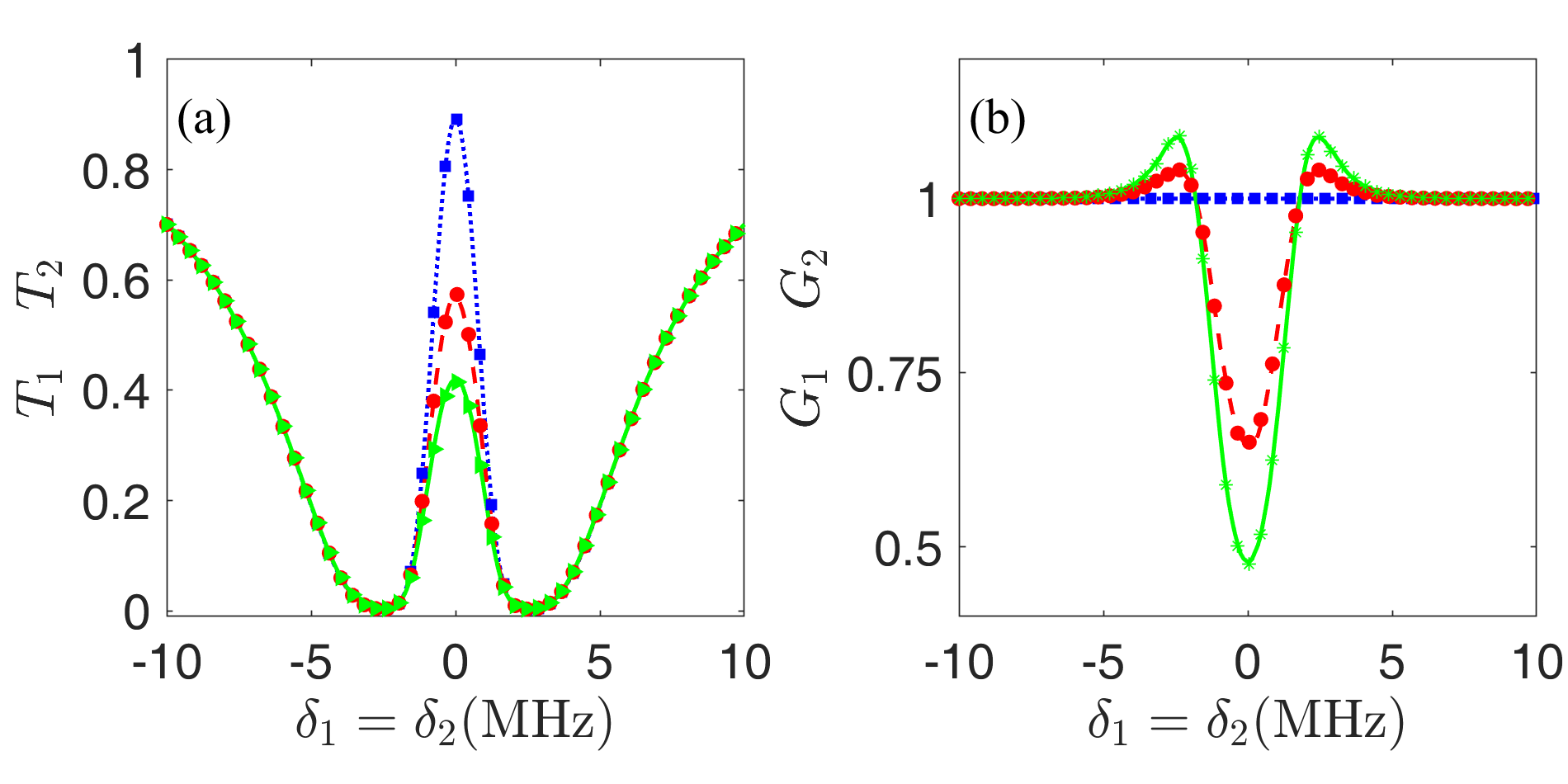}
\caption{(Color online) (a) Probe transmissivities $T_{1}$ and $T_{2}$ and (b) the probe correlations $G_{1}$ and $G_{2}$ versus the
probe detuning $\protect\delta_{1}$ =$\protect\delta_{2}$ for $(\zeta_{1}(0),\zeta_{2}(0))=(0.1, 0.1)$ MHz (blue), $(0.7, 0.7)$
MHz (red), and $(1.0, 1.0)$ MHz (green) with initial two-photon correlation function $g_{1}(0)=g_{2}(0)=1.0$. Lines with and without
symbols represent the first and the second ensembles, respectively. Other parameters are $\Omega _{1}/2\protect\pi=\Omega _{2}/2\protect\pi =2.5$ MHz, $\protect\gamma _{1}/2\protect\pi =\protect\gamma _{2}/2\protect\pi =3.0$ MHz, $\protect\Gamma %
_{1}/2\protect\pi =\protect\Gamma _{2}/2\protect\pi =10.0$ kHz, $C_{6}/2\protect\pi =140\,\text{GHz}\,\protect%
\mu \text{m}^{6}$, $\rho =1.5 \times 10^{8}  \text{mm}^{-3}$, and $L=1.0$ mm.}
\label{fig2}
\end{figure}

The steady optical responses are examined using both the probe transmissivity $T_{a} =I_{a}\left( L\right) /I_{a}\left( 0\right)$ and the probe correlation $G_{a} =g_{a}\left( L\right) /g_{a}\left( 0\right)$ at the exist of each ensemble. In Fig.\ref{fig2}, we present the probe transmissivities $T_{1}$ and $T_{2}$,
and the probe correlations $G_{1}$ and $G_{2}$ by simultaneously varying their input
probe light intensities $\zeta_{1}(0)$=$\zeta_{2}(0)$. It is clear that the optical responses of the two probe fields are identical and exhibit the typical
nonlinearity: on one hand, the stronger the input probe field is, the greater the
absorption within the EIT window; on the other hand, the probe correlation is suppressed within the EIT
window but is enhanced around the Aulter-Townes doublet $\Omega _{1(2)}\approx \pm
2.5$ MHz as the input probe intensity increases.

As a result, the initially classical input fields ($g_{1}(0)=g_{2}(0)=1$) are modified into anti-bunching fields (  $g_{1}(L)=g_{2}(L)<1$) or bunching fields ($g_{1}(L)=g_{2}(L)>1$) between photons by the time they leave the respective
ensemble.

\begin{figure}[tbph]
\includegraphics[width=0.5\textwidth,height=0.4\textwidth]{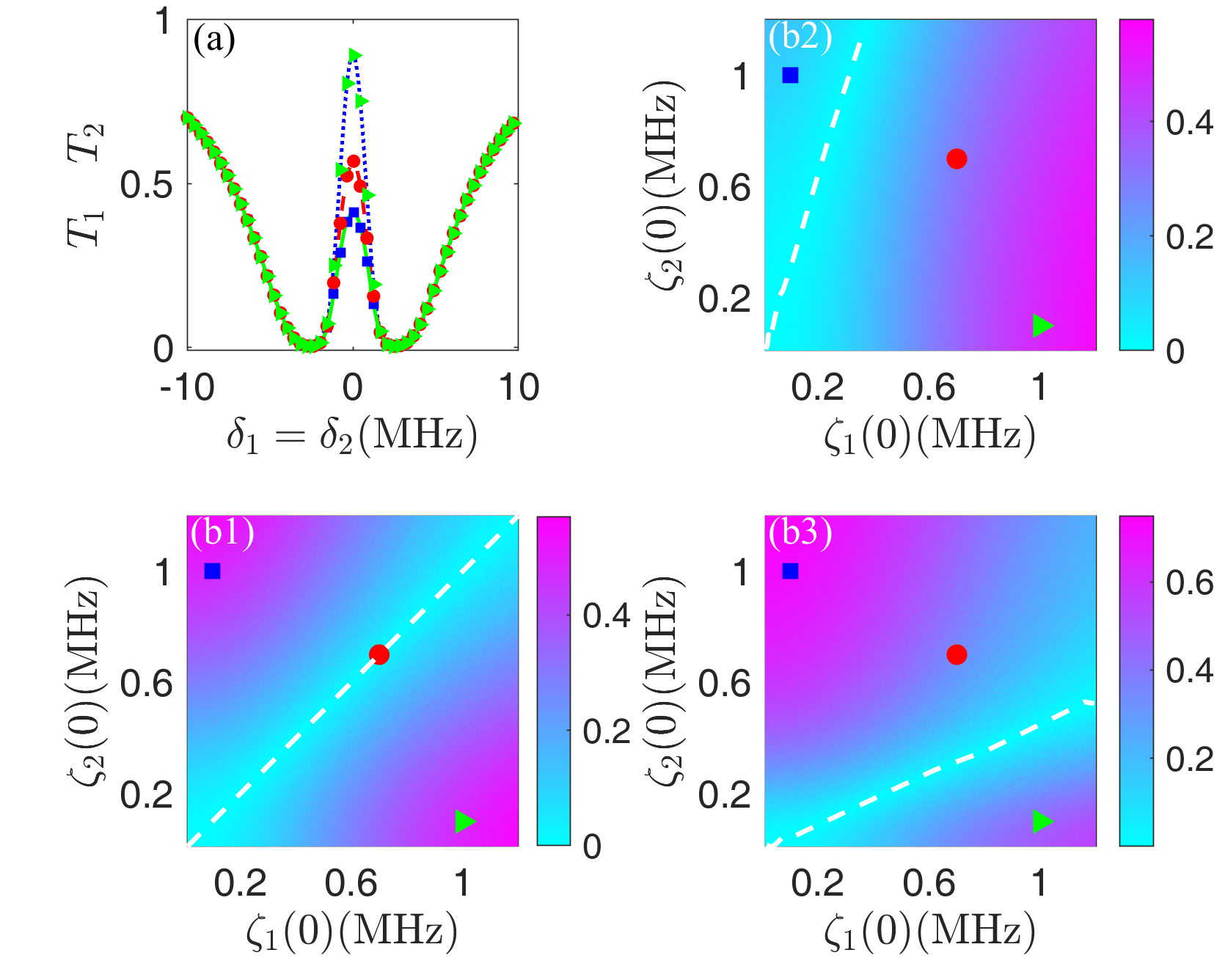}
\caption{(color online). (a) Probe transmissivities $T_{1}$ and $T_{2}$ versus the
probe detuning $\protect\delta_{1}$ =$\protect\delta_{2}$ for $(\zeta_{1}(0),\zeta_{2}(0))=(0.1, 1.0)$ MHz (blue), $(0.7, 0.7)$
MHz (red), and $(1.0, 0.1)$ MHz (green). Lines with and without
symbols correspond to the first and the second ensembles, respectively. (b1)-(b3) Diagrams of the absolute value $|T_2-T_1|$ at $\delta_{1}=\delta_{2}=0$ as a function of the input probe light intensities $\zeta_{1}(0)$ and $\zeta_{2}(0)$ for $g_{1}(0)=1.0$, $0.1$ and $5.0$ with $g_{2}(0)=1.0$. White dashed lines denote $|T_2-T_1|$=0. The blue square, red circle and green triangle mark $(\zeta_{1}(0),\zeta_{2}(0))=(0.1, 1.0)$, $(0.7, 0.7)$
and $(1.0, 0.1)$, respectively. Other parameters are the
same as in Fig.\ref{fig2}.}
\label{fig3}
\end{figure}

Essentially, the symmetry in our system is also evident here. As shown in Fig.~\ref{fig3}(a), $T_{1}$ and $T_{2}$ interchange as $\zeta_{1}(0)$ and $\zeta_{2}(0)$ are
swapped, while all other parameters remain the same. Figure~\ref{fig3}(b1) also displays this symmetry in the parameter space of the input probe light intensities $\zeta_{1}(0)$ and $%
\zeta_{2}(0)$. Figures~\ref{fig3}(b2) and (b3) show that the probe transmissivity is no longer symmetric when the input probe light intensities are exchanged, provided the input two-photon correlation functions differ. For instance, even when $\zeta_{1}(0)= \zeta_{2}(0)$, the optical responses from two
ensembles differ significantly (see the red circles in Figs.~\ref{fig3}(b2) and (b3)). The condition $|T_{1}-T_{2}|\neq 0$ indicates that fewer (or more) photons are absorbed in one ensemble while more (or fewer) photons
are absorbed in the other, due to $g_1(0)\neq g_2(0)$. It is easy to deduce that the symmetry will be broken if different parameters are chosen for the two ensembles, such as the input probe light intensity or the input two-photon correlation function, as mentioned above, as well as the classical control field intensity, atomic density, and other parameters.

\begin{figure}[tbph]
\includegraphics[width=0.5\textwidth,height=0.4\textwidth]{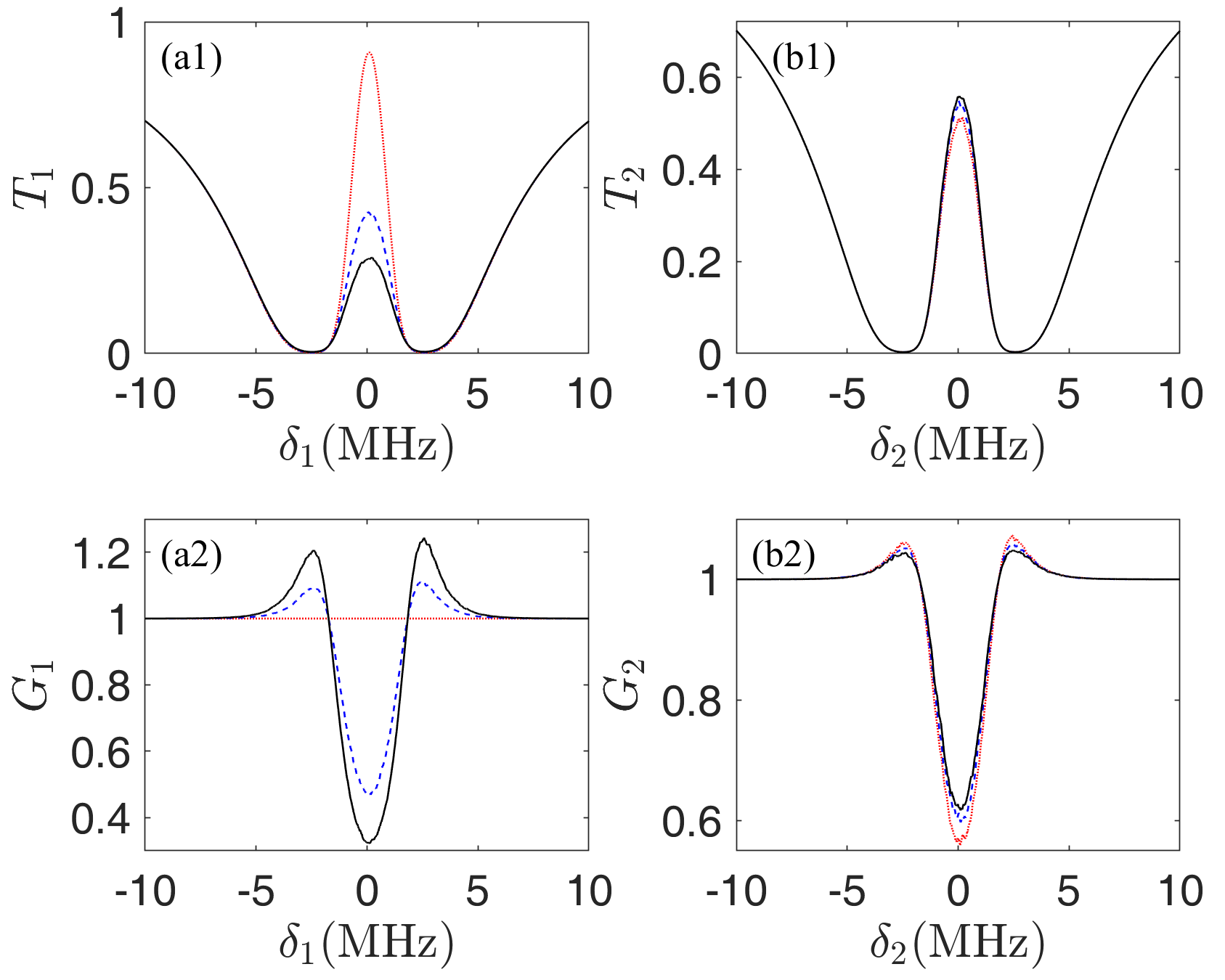}
\caption{(color online). (a1)-(b1) Probe transmissivities $T_{1}$ and $T_{2}$, and (a2)-(b2) the probe correlations $G_{1}$ and $G_{2}$ versus their respective
probe detunings $\protect\delta_{1}$ and $\protect\delta_{2}$ for $\zeta_{1}(0)=0.1$ MHz (red dotted), $0.7$ MHz (blue dashed), and $1.2 $ MHz (black solid) with $\zeta_{2}(0)=1.0$ MHz. Other parameters are the
same as in Fig.\ref{fig2}.}
\label{fig4}
\end{figure}

We then examine the correlations between the optical responses at the two exits of the respective
atomic ensembles. In Fig.\ref{fig4}, we present the probe transmissivities $T_{1}$ and $T_{2}$ and the probe correlations $G_{1}$ and $G_{2}$ by varying $\zeta_{1}(0)$ while keeping $\zeta_{2}(0)$ fixed.  As shown in Figs.\ref{fig4} (a1) and (a2), the first ensemble exhibits clearly optical nonlinearity as $\zeta_{1}(0)$ increases from $0.1$ MHz to $1.2$ MHz. Generally, the optical responses from the two atomic ensembles do not influence each other when the ensembles are completely separated (see the upper schematic diagram in Fig.\ref{fig1}). However, As shown in Figs.\ref{fig4} (b1) and (b2) $T_{2}$ is enhanced by about $10\%$ at $\delta_{2}=0$ without any parameters changes in the second ensemble. Correspondingly, $G_2$ is enhanced by about$12\%$ within the EIT window and is suppressed by $8\%$ around the Aulter-Townes doublet $\Omega _{1(2)}\approx \pm
2.5$ MHz. This correlated phenomenon arises from the complex competition for excitation to the Rydberg state. Specifically, increasing the first input probe field may significantly enhance the probability of excitation to Rydberg state for atoms in the first ensemble and slightly reduce if for atoms in the second ensemble, as they share at most one Rydberg excitation within the same blockade region (a rugby-shaped area of the lower schematic diagram in Fig.1).

\begin{figure}[tbph]
\includegraphics[width=0.5\textwidth,height=0.4\textwidth]{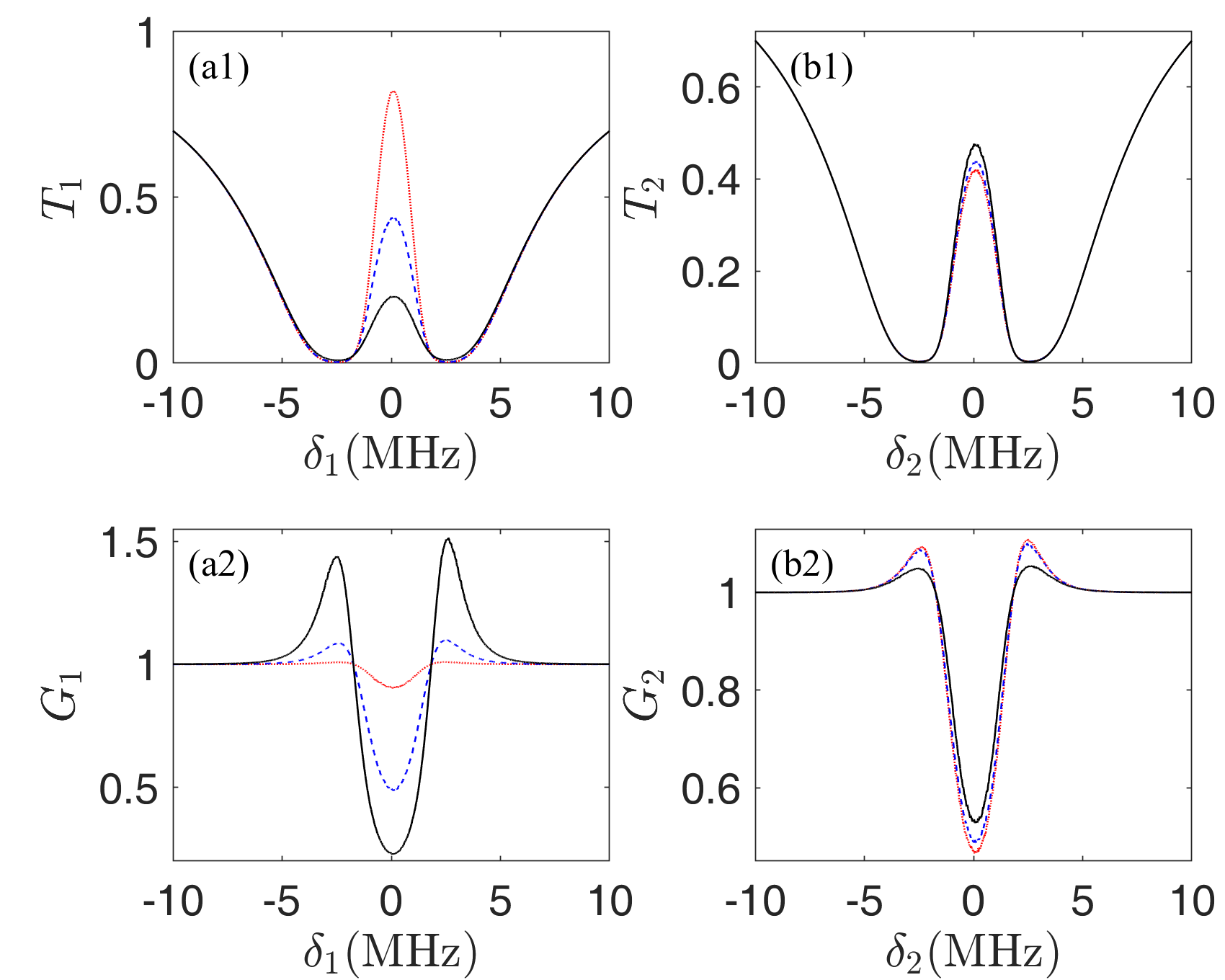}
\caption{(color online). (a1)-(b1) Probe transmissivities $T_{1}$ and $T_{2}$, and (a2)-(b2) the probe correlations $G_{1}$ and $G_{2}$ versus their respective
probe detunings $\protect\delta_{1}$ and $\protect\delta_{2}$ for $g_{1}(0) =0.1$ (red-dotted), $1.0$ (blue dashed), $5.0$ (black solid) with $g_{2} =1.0$. Other parameters are the
same as in Fig.\ref{fig2}.}
\label{fig5}
\end{figure}

In addition, changing the input two-photon correlation function in one ensemble also affects the output optical responses of the other ensemble. Figure.\ref{fig5} shows at $\delta_{2}=0$ both $T_{2}$ and $G_{2}$ are enhanced by about $12\%$, while around the Aulter-Townes doublet $G_2$ is suppressed by approximately $12\%$. This occurs because a bigger two-photon correlation function in the first ensemble increases the probability that the ground-state atoms in the first ensemble will absorb photons. Consequently, the probability of excitation to Rydberg state for atoms in the second ensemble is reduced. Physically, this behavior results from nonlinear optical responses mediated by the Rydberg interaction.

\begin{figure}[tbph]
\includegraphics[width=0.5\textwidth,height=0.206\textwidth]{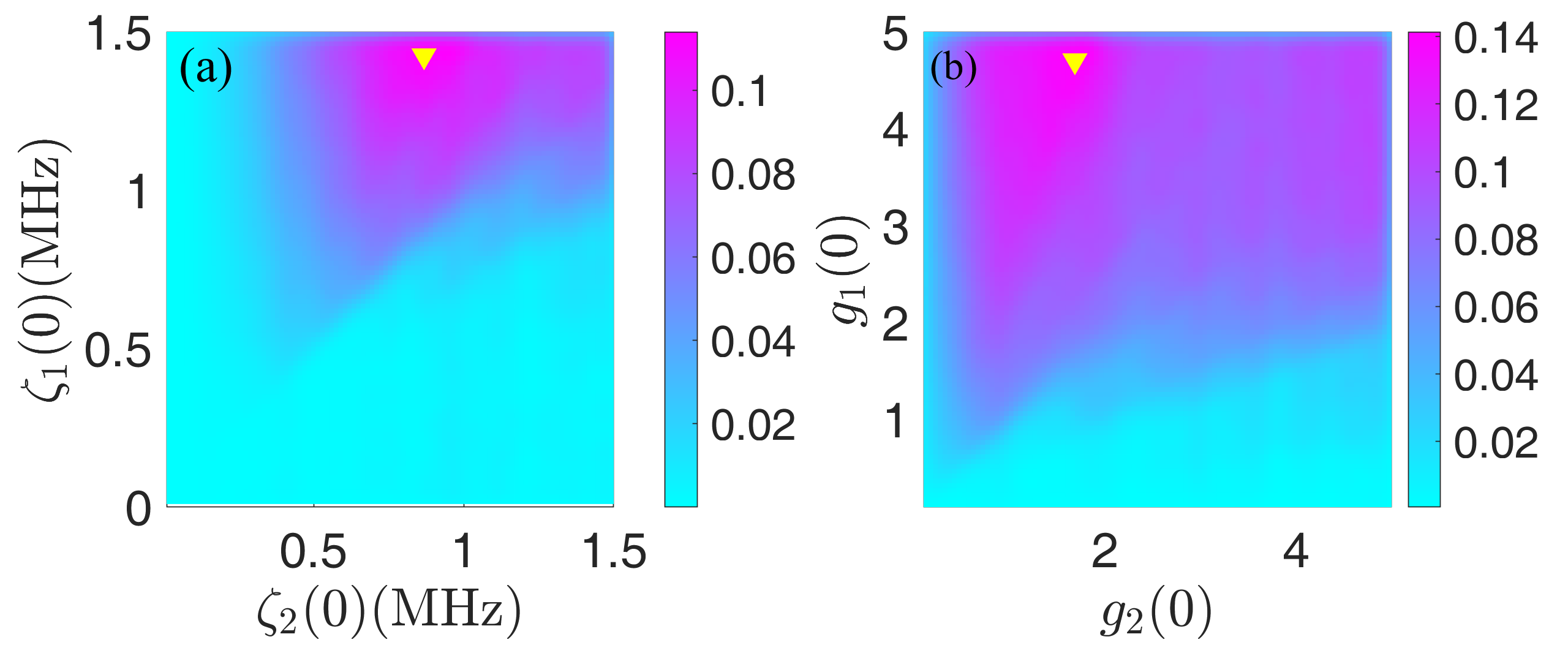}
\caption{(color online). (a) Diagram of the growth rate of the second probe transmissivities $\eta(\zeta_{1}(0),\zeta_{2}(0))=T_{2}(\zeta_{1}(0),\zeta_{2}(0))/T_{2}(\zeta_{1}(0),\zeta_{2}^{0}(0))-1$ with $\zeta_{2}^{0}(0)\equiv 0.01$ MHz as a function of the input probe light intensities $\zeta_{1}(0)$ and $\zeta_{2}(0)$ for $g_{1}(0)=g_{2}(0)=1.0$. (b) Diagram of the growth rate of the second probe transmissivities $\eta(g_{1}(0),g_{2}(0))=T_{2}(g_{1}(0),g_{2}(0))/T_{2}(g_{1}(0),g_{2}^{0}(0))-1$ with $g_{2}^{0}(0)\equiv 1.0$ as a function of the input two-photon correlation functions $g_{1}(0)$ and $g_{2}(0)$ for $\zeta_{1}(0)=\zeta_{2}(0)=1.0$ MHz. The yellow triangle denotes the maximal growth rate. Other parameters are the
same as in Fig.\ref{fig2}.}
\label{fig6}
\end{figure}

Finally, to demonstrate the ability to manipulate the output optical response from one ensemble by adjusting the input optical parameters of the other, we plot the growth rate of the second probe transmissivities by varying the input probe light intensities in Fig.~\ref{fig6}(a) and the input probe correlation function in Fig.~\ref{fig6}(b). Figure~\ref{fig6}(a) shows that $\eta$ is nearly zero when $\zeta_{1}\leq 0.4$ MHz or $\zeta_{2}\leq 0.4$ MHz. In this regime, manipulation is not possible because the input probe light intensities are too weak to excite the Rydberg state in either ensembles, preventing an effective correlation between them. Beyond this threshold, manipulation becomes feasible as the Rydberg interaction between the two ensembles takes effect. $\eta$ always increases with increasing $\zeta_{1}$, but initially increases and then decreases with decreasing $\zeta_{2}$. The maximal growth rate of $14\%$ is achieved at $(\zeta_{2}$, $\zeta_{1})=(0.8,1.0)$ MHz.

Figure~\ref{fig6}(b) shows that the growth rate $\eta$ is nearly zero when $g_{1}\leq 0.6$ or $g_{2}\leq 0.2$. In this case, although the input probe light intensity is strong enough, this light with $g_2\ll 1$ provides too few photons to excite the Rydberg state within a given volume. Therefore, manipulation is ineffective because competition for excitation to Rydberg state between the two ensembles can not occur. Similar to the input probe light intensities, $\eta$ consistently increases with increasing $g_{1}$, and initially increases with increasing $g_2$ until $g_2=0.8$, after which it decreases. The maximal growth rate of $14\%$ occurs at $(g_{2}, g_{1})=(0.8,1.0)$. Overall, the most effective manipulation is achieved when the primary optical parameters exceed the secondary ones, i.e.,  $\zeta_{2}>\zeta_{1}$ and $g_{2}>g_{1}$.

\section{Conclusions}
In summary, we have investigated the correlated steady-state optical responses between two probe fields passing through closely spaced, parallel one-dimensional samples of cold Rydberg atoms. Under the condition of optical nonlinearity, the EIT spectrum of one ensemble can be modified by varying the input probe intensity and the two-photon correlation function of the other ensemble. This capability enables us to perform quantum manipulation with Rydberg ensembles. Furthermore, we systematically investigate the effectiveness of this quantum manipulation. This model can be expanded to multiple Rydberg ensembles to build quantum network and explore quantum work.

\begin{acknowledgements}

This work is supported by the National Natural Science
Foundation of China (Grant Nos. 11874004, 1124019, 12204137,12404299) and the Hainan Provincial Natural Science Foundation of
China (Grant No. 122QN302). This
project is also supported by the specific research fund of The Innovation
Platform for Academicians of Hainan Province (Grant Nos. YSPTZX202215,
YSPTZX202207). 
\end{acknowledgements}

\end{document}